\documentclass[iop]{emulateapj}
\usepackage{amsmath}
\RequirePackage{lineno}
\usepackage{graphicx}
\graphicspath{{Img/}}

\usepackage{enumerate}
\usepackage{tabularx}
\shorttitle{Thermal emission in ultra-long GRB 130925A} \shortauthors{Basak et al.}
\begin{document}

\title{THERMAL EMISSIONS SPANNING THE PROMPT AND THE AFTERGLOW PHASE OF THE ULTRA-LONG GRB 130925A}

\author{Rupal Basak$^{\rm 1,2}$ and A.R. Rao$^{\rm 2}$}

\affil{$^{\rm 1}$ Nicolaus Copernicus Astronomical Center, ul. Bartycka 18, 00-716 Warsaw, Poland \newline
$^{\rm 2}$Tata Institute of Fundamental Research, Mumbai - 400005, India. \newline
$rupal@camk.edu.pl$, $arrao@tifr.res.in$}

\begin{abstract}
GRB~130925A is an ultra-long GRB, and it shows clear evidences for a thermal emission 
in the soft X-ray data of \emph{Swift}/XRT ($\sim0.5$\,keV), lasting till the X-ray 
afterglow phase. Due to the long duration of the GRB, the burst could be studied in hard X-rays with 
high-resolution focusing detectors (\emph{NuSTAR}). The blackbody temperature, as measured
by the \emph{Swift}/XRT, shows a decreasing trend till the late phase (Piro et al. 2014) whereas 
 the high-energy data reveals a significant 
blackbody component during the late epochs at an order of magnitude higher temperature ($\sim5$\,keV),
as compared to the contemporaneous low energy data (Bellm et al. 2014). 
We resolve this apparent contradiction 
by demonstrating that a model with two black bodies and a power-law (2BBPL) is consistent with the data 
right from the late prompt emission to the afterglow phase. 
Both the blackbodies show a similar cooling 
behaviour upto the late time.
We invoke   a structured jet, having a fast spine and a slower sheath layer, to 
identify the location of these blackbodies.
 Independent 
of the physical interpretation, we propose that the 2BBPL model is a generic feature of the prompt 
emission of all long GRBs, and the thermal emission found in the afterglow phase of different GRBs reflects 
the lingering thermal component of the prompt emission with diverse time-scales. 
We strengthen this proposal by pointing out a close similarity between the spectral evolutions 
of this GRB and GRB~090618, a source with significant wide band data during the early afterglow phase. 
 
\end{abstract}


\section{Introduction}

The X-ray and gamma-ray emissions from gamma-ray bursts (GRBs) show two distinct 
phases: a prompt emission with drastic spectral and temporal variations, and an afterglow 
having smooth variation with time and energy (\citealt{Piran_1999_review}). 
Though the afterglow emission is understood as due to synchrotron emission
from a fast moving jet in the ambient medium, the characteristics and the radiation
mechanisms of the prompt emission is still a matter of debate (\citealt{Meszaros_2006}). 
This apparent dichotomy of observations, and the corresponding understanding, is accentuated by 
the fact that the prompt emission is studied using wide band hard X-ray/gamma-ray detectors with 
poor energy resolution and low sensitivity (due to high background), while the 
leisurely X-ray afterglow is amenable to be observed with focusing X-ray telescopes with excellent
energy resolution and sensitivity. The fast moving \emph{Swift} satellite (\citealt{Gehrelsetal_2004_swift}) 
provides good quality data for the early afterglow of several GRBs in the low energies 
with the X-Ray Telescope (XRT; \citealt{Burrowsetal_2005_XRT}). However, the typical slewing ability 
($\sim60 - 80$\,s) is insufficient to capture the prompt emission of the majority of GRBs with the XRT. 
Further, during the XRT observations, the hard X-ray emission fades below the detection level of 
\emph{Swift}/Burst Alert Telescope (BAT; \citealt{Barthelmyetal_2005_swift}) for most cases. 
Hence, it is difficult to identify a broadband spectral feature with the combined BAT-XRT data.

In a recent paper (\citealt{Basak_Rao_2014_090618}, hereafter Paper I), we found a clear evidence for two smoothly 
evolving blackbodies in the early afterglow of GRB~090618. As this GRB has a long gamma-ray duration, a 
significant overlap is seen in the BAT and XRT observations during the junction of the prompt and the 
afterglow phase ($125-165$\,s). This overlapping observation 
allowed us to identify two blackbodies in the individual energy windows provided by the BAT and the XRT. The temperature 
and the flux of the two blackbodies are found to have similar evolution throughout the final pulse of the prompt 
emission till the early afterglow phase. This is one of the rare GRBs with a significant
overlap in the XRT and BAT observations, and further, the two blackbodies happen to 
independently dominate the two detectors. Hence it is important to confirm this finding and examine whether
this behaviour is a generic feature of all GRBs. 

In this context ultra-long GRBs, a class of GRBs which produce prompt emission for hours, provide an
excellent opportunity to study the spectral evolution from the prompt to the afterglow phase.
GRB~130925A is an ultra-long GRB with the gamma-ray duration of $\sim2$\,hr. This GRB was detected by several 
satellites including the \emph{Swift}/BAT and the \emph{Swift}/XRT. Due to the gamma-ray emission on a long time-scale, 
the burst could be observed with the focusing telescopes of \emph{NuSTAR} (\citealt{Harrisonetal_2013}).
The \emph{Swift}/XRT data revealed the presence of a blackbody emission at low energies
during the prompt to the afterglow phase (\citealt{Piroetal_2014}; P14 hereafter), though the same data 
could be interpreted as due to a steep power-law emission rather than a blackbody (\citealt{Evansetal_2014}).
The hard X-ray data from the \emph{NuSTAR}, on the other hand, found a very significant emission
above $\sim$10 keV at late times, which could be interpreted as due to an emission from a blackbody of 
temperature $\sim$5 keV (\citealt{nustar_2014}; B14 hereafter). The contemporaneous data 
of \emph{Swift}/XRT, however, finds a blackbody with temperature $\sim$0.5 keV (P14).
In this \emph{Letter} we demonstrate that all 
these data are consistent with a model consisting of two black bodies and a power-law (2BBPL). 
In the next section we summarise the pertinent data on GRB~130925A
and present a re-analysis. The results are discussed in the context of a spine-sheath jet
(section 3), and in the last section we give a summary of the important findings of this work.

\section{Thermal Emission in GRB 130925A}

GRB~130925A was triggered by the \emph{Swift}/BAT at $T_0=$ 2013-09-25 04:11:24 UT. The \emph{Swift}/XRT
followed the burst from 150\,s after the BAT trigger. The burst showed several pulse emissions and 
soft flares till late time. A detailed analysis of the high-energy emission from this source 
is presented in \cite{Evansetal_2014}. They argue that this GRB (and by extension other ultra-long GRBs) 
could be understood as events with low circumburst densities, and correspondingly having an order of magnitude 
lower deceleration time compared to the normal long GRBs. They strongly favour a steep power-law function 
(index $\Gamma\sim-4$) for the late X-ray afterglow spectrum, and use the dust scattering model for the explanation
(\citealt{Shao_Dai_2007}). 

P14, on the other hand, explain the steep spectrum as a superposition of a blackbody (temperature, $kT\sim0.5$\,keV)
and a power-law (BBPL) model. A crucial \textit{XMM-Newton} observation after 3 months  shows a harder power-law 
index ($\sim-2.5$). As the thermal emission possibly subsides after such a long time, the spectrum is expected to reveal 
the underlying non-thermal component. The hard value of the index agrees with a standard afterglow in a wind medium. 
On the other hand, the blackbody component is suggested to be a `hot' cocoon which forms on the GRB jet as it pierces 
through the progenitor --- possibly a blue suergiant star. 

Another important clue on the spectral shape is provided by the observation with high-resolution focusing 
detectors, \emph{NuSTAR}, in three periods (1.8\,day, 8.8\,day, and 11.3\,day), and \emph{Chandra} ACIS-S in one period (11.0\,day). 
B14 have analyzed the spectra in two epochs --- $\sim2$\,day and $\sim9-11$\,day. In both cases, 
they have found significant deviation from the canonical afterglow spectrum. The high-resolution 
data shows a `dip' in the continuum spectrum. B14 have used a BBPL model to capture the feature. Interestingly, they have 
found that the blackbody temperatures in the two epochs ($5.6^{+2.1}_{-1.2}$\,keV and $4.0^{+0.7}_{-0.6}$\,keV, 
respectively) have an order of magnitude higher values than that found by P14 in the XRT data. Though the data 
allows a softer blackbody (near $\sim0.5$\,keV), B14 have argued in favour of the higher-temperature blackbody, 
as the lower-temperature blackbody shows an unphysical contraction of the apparent emitting region at the later epoch, from $(3.2\pm0.8)\times10^{10}$\,cm to $(1.7\pm0.4)\times10^{10}$\,cm.

\begin{table}
\caption{Parameters of the two blackbodies in the two \emph{NuSTAR} observation}

\vspace{0.1in}

\begin{tabular}{c|ccc}
\hline
Interval & $kT$ & $F_{\rm BB}$ & $R_{\rm BB}$ \\
& (keV) & (erg\,cm$^{-2}$\,s$^{-1}$) & $10^8$\,cm \\
\hline
\hline
\multicolumn{4}{l}{Observation 1 ($\sim T_0+1.8$ day)}\\
\hline
$^{(a)}$BB$_1$  & $5.12_{-1.52}^{+2.79}$ & $(5.46\pm2.58)\times10^{-13}$ & $0.9\pm0.3$\\
BB$_2$ & $0.42_{-0.10}^{+0.11}$ & $(1.25\pm0.92)\times10^{-12}$ & $(2.0\pm1.0)\times10^2$\\
\hline
\multicolumn{4}{l}{Observation 2 ($\sim T_0+11$ day)}\\
\hline
BB$_1$  & $3.0_{-0.5}^{+0.6}$ & $(1.3\pm0.3)\times10^{-13}$ & $1.3\pm0.2$\\
BB$_2$ & $0.15\pm0.03$ & $(2.0\pm1.4)\times10^{-12}$ & $(2.0\pm1.0)\times10^3$\\
\hline

\end{tabular}

\vspace{0.1in}
\begin{footnotesize}
$^{(a)}$BB$_1$=Higher-temperature blackbody, BB$_2$=Lower-temperature blackbody
\end{footnotesize}

\label{t1}

\end{table}

\begin{table*}\centering
\caption{Parameters of the BBPL model fitted to the BAT data in 105-150\,s interval, 
and that of the 2BBPL model fitted to the joint BAT-XRT data in 150-295\,s interval}

\vspace{0.1in}

\begin{tabular}{ccccc}
\hline
Parameter & 105-150\,s & 150-175\,s & 175-225\,s & 225-295\,s \\
\hline
\hline
$n_{\rm H}$ ($10^{22}$\,atoms\,cm$^{-2}$) & & $2.1\pm0.2$     & $1.7\pm0.2$                 &  $1.7\pm0.2$ \\

$kT$\,(keV)  & $13.0_{-3.7}^{+4.0}$                  & & & \\ 
$N$          & $0.11_{-0.06}^{+0.09}$                   & & & \\

$kT_h^{(a)}$\,(keV)            &  & $8.9_{-1.7}^{+1.8}$      & $7.0_{-3.5}^{+5.8}$         & $8.5_{-7.1}^{+4.6}$  \\ 
$N_h$                          &  & $0.11_{-0.04}^{+0.05}$   & $(3.5\pm3.2)\times10^{-2}$  & $(3.2_{-2.5}^{+3.2})\times10^{-2}$ \\
$kT_l$\,(keV)                  &  & $2.1_{-0.6}^{+0.5}$      & $2.1\pm0.5$                 & $2.3_{-0.7}^{+0.5}$ \\
$N_l$                          &  & $0.14_{-0.06}^{+0.07}$   & $(7.6_{-3.5}^{+4.6})\times10^{-2}$  & $(7.7_{-7.4}^{+12.0})\times10^{-2}$\\
$\Gamma$                       &  $-2.3\pm0.2$ & $-2.2\pm0.1$             & $-2.1\pm0.1$                & $-2.1\pm0.1$ \\
\hline
$\chi^2_{\rm red}$\,(dof)      & 0.93 (54) & 1.2 (76) & 1.3 (76) & 1.1 (76)\\
\hline
\end{tabular}

\vspace{0.1in}

\begin{footnotesize}
$^{(a)}$ $h$=Higher-temperature blackbody, $l$=Lower-temperature blackbody
\end{footnotesize}

\label{t2}

\end{table*}

\subsection{Signature of two blackbodies in the afterglow data}
\begin{figure}
{

\includegraphics[width=3.2in]{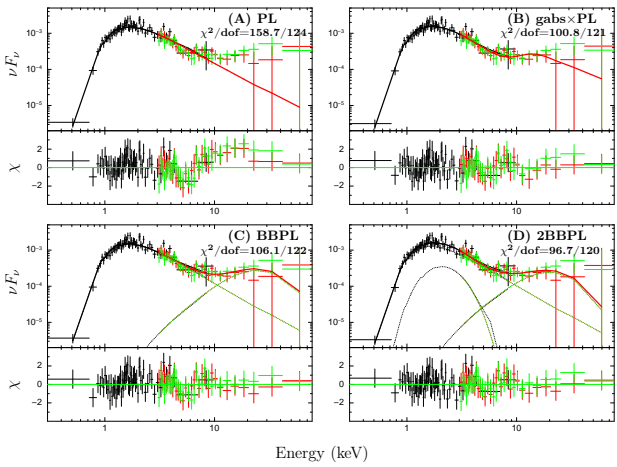} 

}
\caption{Spectral fitting to the combined data of \emph{NuSTAR} and \emph{Swift}/XRT of GRB 130925A during 
1.8 day after the \emph{Swift}/BAT trigger (see text for explanation).
}
\label{fig1}
\end{figure}

To arrive at a correct spectral model, it is important to fix the continuum and hence data above $\sim$10 keV is very crucial. 
We re-analyze the data whenever a higher energy observation is available (either BAT or \emph{NuSTAR} observation). 
These epochs are: (I) $150-300$\,s during the prompt emission (joint BAT-XRT observations), 
(II) 1.8 day during the afterglow (joint \emph{NuSTAR}-XRT observations), and (III) 
$\sim9-11$ day (joint \emph{Chandra}-\emph{NuSTAR} observations). 
The observation in epoch II provides the best possible data among all the observations, 
due to the superior sensitivity and resolution of the \emph{NuSTAR} detectors (and
the higher flux compared to epoch III).
In Figure~\ref{fig1}, we show the spectrum of epoch II, fitted with various models. We follow the standard procedure 
described by B14 to extract the spectral data. The spectrum is fitted in {\tt XSPEC v12.8.2}. We fit the following models:
(A) power-law (PL), (B) Power-law with a Gaussian absorption (gabs$\times$PL), (C) a blackbody with a power-law (BBPL), and 
(D) two blackbodies with a power-law (2BBPL). In all cases we have used a {\tt wabs} and a {\tt zwabs} model to account for 
the absorption in our Galaxy and in the source, respectively. 
P14 have fitted the time-resolved data during 20\,ks-8\,Ms with 
{\tt wabs}$\times${\tt zwabs}$\times${\tt PL} model. As the equivalent hydrogen column density ($N_{\rm H}$) at the source should 
not change over time, they have linked this parameter in all time bins to obtain its precise value --- 
$(2.1\pm0.1)\times10^{22}$\,cm$^{-2}$. We have used this value, along with the Galactic absorption of  
$1.7\times10^{20}$\,cm$^{-2}$ to perform uniform fitting with all the 
models. Keeping the source absorption free gives a value close to P14, within the statistical errors. 
Note in Figure~\ref{fig1} (panel A) that a PL fit shows a `dip' near 5\,keV. We obtain $\chi^2$/ degrees of freedom 
(dof)=158.7/124 for a PL fit. Incidentally, a Band function (\citealt{Bandetal_1993}) gives a similar unacceptable fit with 
$\chi^2$/dof=153.0/121. A Gaussian 
absorber gives an immediate improvement with $\chi^2$/dof=100.8/121. However, as noted by B14, we also see a shift in the 
centroid from $6.4\pm0.8$\,keV to $<4.0$ keV from epoch II to epoch III. Hence, the Gaussian absorber is unphysical. A BBPL 
model gives a reasonable fit with $\chi^2$/dof=106.1/122 (see panel C). Finally, we fit the spectrum with a 2BBPL model and obtain 
$\chi^2$/dof=96.7/120. The parameters of the blackbodies are shown in the first two rows of Table~\ref{t1}. 
We perform F-test to find the significance of adding the second blackbody. We find a $2.9\sigma$ significance
corresponding to $p-{\rm value}=3.8\times10^{-3}$. The temperature of the two blackbodies 
in this epoch are: $5.12_{-1.52}^{+2.79}$\,keV and $0.42_{-0.10}^{+0.11}$\,keV. Note that 
these values are close to those reported by B14 and P14 (both within $1\sigma$), respectively. 
Based on the analysis of the epoch II, we fit the spectrum of the epoch III with a 2BBPL model. We freeze the power-law 
index to the value obtained in epoch II (-3.74). B14 has combined the \emph{NuSTAR} data of $T_0+8.8$ day 
and $T_0+11.3$ day. As both the blackbodies may evolve significantly during this time, we use only the data 
of $T_0+11.3$ day. We obtain a $\chi^2$/dof=146.55/133. The parameters of the blackbodies are shown in the last 
two rows of Table~\ref{t1}. The temperature of the two blackbodies are $3.0_{-0.5}^{+0.6}$\,keV 
and $0.15\pm0.03$\,keV. We note that the temperature of the higher-temperature blackbody is somewhat lower than the temperature 
of the single blackbody as obtained by B14 ($4.0_{-0.6}^{+0.7}$).

\subsection{Thermal emission spanning the prompt and afterglow phase}
In order to see how the temperature of the two blackbodies evolve during the prompt emission, we fit the joint BAT-XRT data in 
150-300\,s interval. We obtain three time-resolved spectra in 150-175\,s, 175-225\,s, and 225-295\,s intervals. For each case,
we obtain a reasonable fit with reduced $\chi^2$ close to 1 for 76 dof (see Table~\ref{t2}). In the first interval, the 2BBPL model shows significant 
improvement compared to a BBPL model with $\triangle \chi^2=18.6$ in the expense of two dof. The F-test gives a $3.3\sigma$ significance 
($p-{\rm value}=9.3\times10^{-4}$) for the addition of the second blackbody. It is worthwhile to mention that a Band 
function also gives an acceptable fit to the prompt emission data. However, as already mentioned
the Band function is inadequate during the epoch II, where we have good quality data. In addition to the joint BAT-XRT data,
we also fit a BBPL model to the falling part of the final pulse (105-150\,s) and obtain $kT=12.9_{-3.7}^{+4.0}$\,keV. 
 
\begin{figure}
{

\includegraphics[width=3.2in]{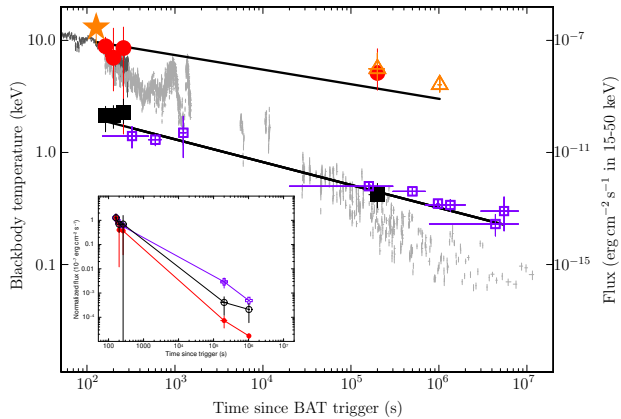} 

}
\caption{Evolution of the temperature of the two blackbodies found in various epochs of GRB~130925A 
(filled symbols for present work, star for the falling part of the last pulse --- 105-150\,s).
The data from P14 and B14 are shown by open squares, and open triangles, respectively. 
The 15-50 keV flux is shown in the background. \emph{Inset:} Flux evolution of different components --- 
filled circles: higher-temperature blackbody, open circles: lower-temperature blackbody, plus: power-law. 
The flux of the two blackbodies are normalized with the power-law flux at the initial time bin 
(multiplied by 13.6 and 10.7, respectively). 
}
\label{fig2}
\end{figure}

In Figure~\ref{fig2}, we show the temperature evolution of the two blackbodies (filled symbols --- circles for the higher-, 
and squares for the lower-temperature blackbody). The lightcurve (15-50\,keV flux in units of erg\,cm$^{-2}$\,s$^{-1}$) 
of the GRB in the BAT and XRT detectors is shown in the background. The star represents the single blackbody temperature during the 
falling part of the last pulse in the BAT data. For a comparison, we have plotted the temperature values obtained by P14 
(open squares) and B14 (open triangles). We note that the values of the higher and the lower temperature are in general 
agreement with that of B14 and P14, respectively. We fit the evolution of both the blackbodies as a power-law function 
of time. In doing this, we assume that all the blackbody temperatures given by P14 are that of the lower-temperature blackbody of the unified 
2BBPL model. The slope of the evolution are $-0.15\pm0.05$ and $-0.20\pm0.02$ for the higher- and lower-temperature blackbody, respectively. 
The two evolutions can be considered either approximately similar, or the higher-temperature blackbody has a slower temperature evolution. 
Note that due to the large gap in the prompt and afterglow coverage of the high energy observation, the error in the slope 
of the higher-temperature blackbody evolution is large. Hence, the comparison between the two temperature evolutions remains inconclusive.
In the inset of Figure~\ref{fig2}, we show the time evolution of the flux of the individual components of the 2BBPL model. 
We have shown only those points, where we have high energy observation with either BAT or \emph{NuSTAR}. In order to compare 
the flux evolution of the components, the blackbody flux are normalized to the power-law flux of the initial time bin. 
We note that the thermal flux decreases more rapidly compared to the power-law flux. The flux of the higher-temperature blackbody evolves 
even faster. 

\section{Discussion}

Though it is not conclusive that the 2BBPL is the correct description of
the GRB emission model during the prompt and the early afterglow phase, there are 
strong evidences to support the claim that this model is consistent with a wide variety
of data. This model is 
superior to other models (BBPL/ Band) for bright GRBs (\citealt{Raoetal_2014}) and it
is a preferred model for GRBs with single pulses 
(\citealt{Basak_Rao_2014_MNRAS}), GRBs with separable pulses 
(\citealt{Basak_Rao_2013_parametrized}) and GRBs with high energy 
(GeV) emission and detected by \emph{Fermi}/Large Area Telescope (\citealt{Basak_Rao_2013_linger}).
For GRB 090618 (Paper I), the superior spectroscopic \emph{Swift}/XRT detector
shows a blackbody temperature decreasing with time, consistent with the 
interpretation that the two components of the 2BBPL model seen in the 
last GRB pulse are showing a cooling behaviour. 
 In the present paper, we have 
seen yet another evidence for the two blackbody emissions in the case of GRB 130925A.
In the next sections, we investigate the implications for the putative photospheres
and a possible jet scenario where two photospheres can co-exist.
 
\subsection{Photosphere of the two blackbodies}
It is interesting to calculate the apparent emission radius for a spherical emission region ($R_{\rm BB}$) during the two epochs of 
\emph{NuSTAR} observation. 
Both B14 and P14 find that the value of $R_{\rm BB}$ shows 
a contraction during this time.
 As we have 
two blackbodies, we calculate the value of $R_{\rm BB}$ for each of them as follows.

\begin{equation}
 R_{\rm BB}=\sqrt{\frac{d_{\rm L}^2 F_{\rm BB}}{\sigma [(1+z)T]^4}}
\end{equation}

where $d_{\rm L}$ is the luminosity distance, $\sigma$ is Stefan-Boltzmann constant $=5.6704\times10^{-5}$ 
erg\,cm$^{-2}$\,s$^{-1}$\,K$^{-4}$, $F_{\rm BB}$ is the observed blackbody flux, and $(1+z)T$ is the temperature 
corrected for the redshift, $z$. The radius of the higher-temperature blackbody in the two epochs are found to be 
$R_{\rm BB}=(0.9\pm0.3)\times10^8$\,cm and $R_{\rm BB}=(1.3\pm0.2)\times10^8$\,cm, which show a slow increment. 
Interestingly, the value of $R_{\rm BB}$ of the lower-temperature blackbody is found to be $R_{\rm BB}=(2.0\pm1.0)\times10^{10}$\,cm 
and $R_{\rm BB}=(2.0\pm1.0)\times10^{11}$\,cm (see the last column of Table~\ref{t1}). Note that this result is markedly 
different from that obtained by B14 and P14. 
Our analysis shows a definite increment of the apparent radius at the later 
epoch.

\subsection{A spine-sheath jet}
In paper I, we have proposed a spine-sheath jet to explain the evolution of the two blackbodies. A fast moving spine
with a slower sheath component is expected in a number of physical scenario, e.g., a hot cocoon formed over the 
GRB jet as it pierces through the envelop of the progenitor (\citealt{Meszaros_Rees_2001, Ramirez-Ruizetal_2002, 
Zhangetal_2003, Zhangetal_2004}), or a collimated proton jet along with a wider neutron sheath for a magnetic jet 
(\citealt{Vlahakisetal_2003, Pengetal_2005}). Such a structured jet is frequently invoked to explain double jet 
break and optical re-brightening (e.g., \citealt{Bergeretal_2003_spinesheath, Liang_Dai_2004, Hollandetal_2012}).
If the sheath component of the jet is in fact the cocoon, we expect it to be less collimated, and the terminal 
bulk \emph{Lorentz} factor ($\eta$) to be lower by a factor of $\sim10$ than the spine.


The observed physical sizes of the two blackbodies show 
very slow expansion rates and the corresponding sizes of the 
photospheres too would have slow speeds and hence
they cannot be tied to the actual sizes of the outflowing
plasma. Hence the photospheres have to be steady/ gradually changing
regions in the outflow. We invoke that the spine photosphere produces the higher-temperature blackbody 
while the lower-temperature blackbody is produced by the sheath photosphere.  
 Note that the size of the emitting regions calculated above are considered as spherical emission regions. The physical 
photosphere can be related to the observed size  by $r_{\rm ph}=\eta R_{\rm BB}$. With reasonable values of 
$\eta$ for the spine and the sheath
(say $10^3$ and $10^2$, respectively), one should obtain larger values of the corresponding emission radius. 
The observed properties of the two photospheres can be
used to constrain the properties of the spine and the sheath.
For example, 
  we have 
found that the higher-temperature blackbody has much lower  evolution
and hence we can assume  that the spine remains 
``steady'' for a long time. The sheath photosphere, on the other hand, shows considerable evolution. 
For a cocoon sheath, a rapid photospheric expansion is indeed expected (\citealt{Starlingetal_2012}).

\subsection{Comparison with GRB~090618}

\begin{figure}
{

\includegraphics[width=3.2in]{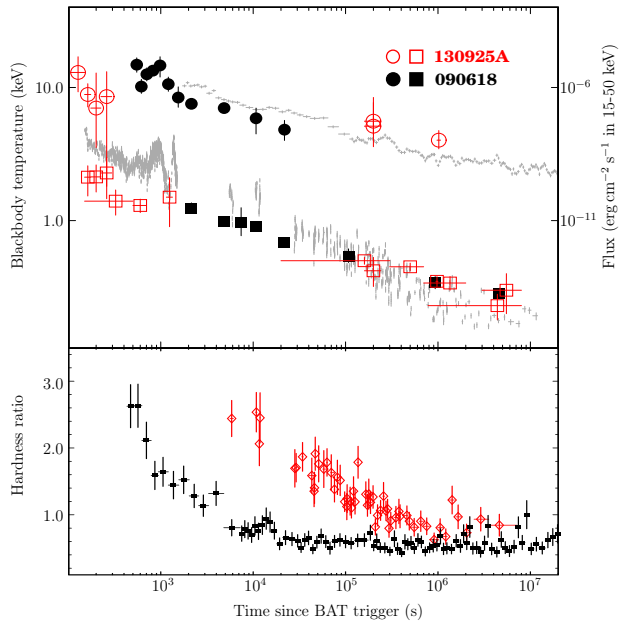} 

}
\caption{A comparison of spectral evolution between the ultra-long GRB~130925A with GRB~090618, a typical long GRB
(open and filled symbols, respectively). The 100-300\,s time axis of the latter is stretched in the interval 100\,s 
to $2\times10^7$\,s. \emph{Upper panel:} Temperature evolution of the two blackbodies. We have shown the 15-50\,keV 
flux in the background (higher flux for GRB~090618). \emph{Lower panel:} Hardness ratio (count ratio in 1.5-10\,keV 
to that of 0.3-1.5\,keV) of the two GRBs. 
}
\label{fig3}
\end{figure}

\cite{Evansetal_2014} have made a systematic analysis of 672 XRT GRBs,
and found that after 3 ks of the individual trigger, the hardness ratio (HR) does not
show any evolution for the majority of GRBs. For 12 GRBs (including GRB~130925A),
which show $>5\sigma$ significance of HR variation at late times, the power-law index 
shows a continuous increment with time. They argue that the HR evolutions in these 
GRBs including GRB~130925A can be explained by a dust scattering model.

However, the hard X-ray data at late times requires emission above 10\,keV, 
which is not compatible with a steep power-law model (as required by the XRT data 
alone). Hence, we explore here whether the spectral evolution of GRB~130925A could 
be explained by the 2BBPL model, and whether this model is a generic feature of all GRBs. 
We have taken the temperature evolution as given in Figure~\ref{fig2}, and verified that 
the count rate variation in two energy bands in the XRT data ($0.3 - 1.5$\,keV and $1.5 - 10$\,keV) 
after 100\,ks is compatible with the flux variation of the three components as given in the 
inset of Figure~\ref{fig2}. 

The rare occurrence of spectral variation after 3\,ks could be explained by 
postulating that the blackbody components decay at different rates for different 
GRBs. We investigate this hypothesis by comparing the results of GRB~130925A
with GRB~090618. In Figure~\ref{fig3}, we plot the relevant parameters for these 
two GRBs, but the time axis of GRB 090618 is arbitrarily stretched ($100 - 300$\,s 
data is stretched to $100- 2\times$10$^7$\,s). The close similarity in the spectral 
variation in these two GRBs encourages us to hypothesise that the 2BBPL model is a generic 
feature of all GRBs and the black body components decay at different rates for different GRBs.

\section{Conclusions}
The major conclusions from the work presented here can be summarized as follows.

(1) We use high-resolution \emph{NuSTAR} data of the ultra-long GRB~130925A, and find that the spectrum is consistent 
with a 2BBPL model. The lower-temperature blackbody is consistent with an evolving blackbody throughout the burst (P14). 
The higher-temperature blackbody is also consistent with the single blackbody of B14, and this component 
also shows a similar evolution as the lower-temperature blackbody.

(2) The evolution of the corresponding photospheres of the two blackbodies are found to be different. While the higher-temperature
blackbody shows a constant spherical emission region, the lower-temperature blackbody shows a rapid evolution of the photosphere.

(3) We have proposed a structured jet with fast spine and slow sheath as the origin of the thermal components. The higher-temperature
blackbody is produced at the spine photosphere, which remains `steady' at the later epoch. The sheath is possibly a cocoon,
which shows a rapid expansion of the photosphere. 

In recent years, we have found that the time-resolved prompt emission spectrum of a variety of GRBs is consistent 
with the 2BBPL model (\citealt{Basak_Rao_2013_parametrized, Basak_Rao_2013_linger, Raoetal_2014}). 
This spectral shape is statistically preferred for GRBs with high signal-to-noise data. However, due to the limited 
spectral resolution of the gamma-ray detectors, we could not draw a firm conclusion. We could detect the two 
blackbodies in the early afterglow data of GRB~090618 (Paper I), as the XRT provided a good spectral data 
during the overlapping observation with the BAT. The current observation also shows a clear 
signature of the two blackbodies which evolve from the very early phase during the prompt emission, and remains 
visible with high significance at very late times. We emphasize again that independent of the physical interpretation of the 
2BBPL, this model emerges as a generic spectrum of long GRBs.


\section*{Acknowledgments} This research has made use of data obtained through the
HEASARC Online Service, provided by the NASA/GSFC, in support of NASA High Energy
Astrophysics Programs. This work made use of data supplied by the UK Swift Science 
Data Centre at the University of Leicester. We thank the referee for valuable comments
to make the presentation more rigorous.

\bibliographystyle{apj}
\bibliography{thesis_bib}

\end{document}